\documentclass[prl,preprint,showpacs,superscriptaddress]{revtex4}
\usepackage{epsfig}

\usepackage{amsmath}
\usepackage{dcolumn}
\usepackage{bm}
\usepackage{color,colordvi,graphicx,pstcol}

\newcommand{\hide}[1]{}

\newcommand{\eq}[1]{Eq.\,(\ref{#1})}

\newcommand{\noeq}[1]{(\ref{#1})}
\newcommand{\fig}[1]{Fig.\,\ref{#1}}

\newcommand{\beq}{\begin{equation}}
\newcommand{\eeq}{\end{equation}}

\begin{document}
\title{Image Storage in Hot Vapors}

\author{L. Zhao}
\affiliation{Department of Physics, University of Connecticut,
Storrs, CT 06269}

\author{T. Wang}
\affiliation{Department of Physics, University of Connecticut,
Storrs,  CT 06269}

\author{Y. Xiao}
\affiliation{Harvard-Smithsonian Center
for Astrophysics, Cambridge, MA 02138}

\author{S. F. Yelin}
\affiliation{Department of Physics, University of Connecticut,
Storrs, CT 06269} \affiliation{ITAMP, Harvard-Smithsonian Center
for Astrophysics, Cambridge, MA 02138}

\date{\today}
\begin{abstract}
We theoretically investigate image propagation and storage in hot atomic vapor. A $4f$ system is adopted for imaging and an atomic vapor cell is placed over the transform plane. The Fraunhofer  diffraction pattern of an object in the object plane can thus be transformed into atomic Raman coherence according to the idea of ``light storage''. We investigate how the stored diffraction pattern evolves under diffusion. Our result indicates, under appropriate conditions, that an image can  be reconstructed with high fidelity. The main reason for this procedure to work is the fact that diffusion of opposite-phase components of the diffraction pattern interfere destructively. 
\end{abstract}


\maketitle
Manipulating images all-optically may play a significant role in many fields, including holography, remote sensing, classic or quantum correlations, image and information processing, and so on. In many applications, the amplitude and phase of images should be preserved. Recently, some studies of the propagation and storage of transverse images were performed \cite{im1,im2,im3,im4}. 
Electromagnetically induced transparency (EIT), a phenomenon of quantum interference, has been investigated for almost two decades \cite{eit1,eit2,eit3}. 
In the EIT system, the propagation of light fields can be described by the coupled light-matter excitation termed ``dark-state polariton''. 
The weak probe light can be manipulated coherently and all-optically, and its amplitude and phase can be preserved \cite{ls1,ls2,ls3,ls4,ls5}. Hence, EIT system may be a good candidate for manipulating images \cite{im5}.

To date, the EIT system has been widely studied in many media. Hot atomic vapor cells are the work horse among those because of ease of use and fabrication \cite{hot1,hot2}. In the vapor, however, the Raman coherence carried by the atoms will diffuse due to the motion in the hot gas \cite{diff1,diff2}. Therefore, the transverse distortion induced by diffusion could pose a real challenge to processing images, especially small ones.

        In this paper, we theoretically demonstrate that, 
similar to the storage of optical vortices \cite{diff2}, the dark spots of the Fraunhofer diffraction pattern of an object in a $4f$ imaging system can be stored for a long time under strong diffusion conditions. The essence of such stability depends only on the destructive interference of Raman coherence. 
Furthermore, using the principle of Fourier optics \cite{fourier}, the spatial information of the object can be mapped into the diffraction pattern. 
Under appropriate conditions, an image with high fidelity 
can form on the screen in the presence of diffusion. 
In our scheme, the $4f$ system is a natural spatial spectrum transformer to produce the Fraunhofer diffraction patten. The destructive interference between different parts can automatically make the stored pattern robust under diffusion. In contrast to the recent image storage \cite{im3}, our scheme is free from the artificial ``phase-shift lithography'' technique  \cite{psl}. (In practice, for an unknown arbitrary image it might be difficult to make the corresponding phase plate beforehand.)

\begin{figure}[ht]
   \centerline{ \includegraphics[clip,width=1.0\linewidth]{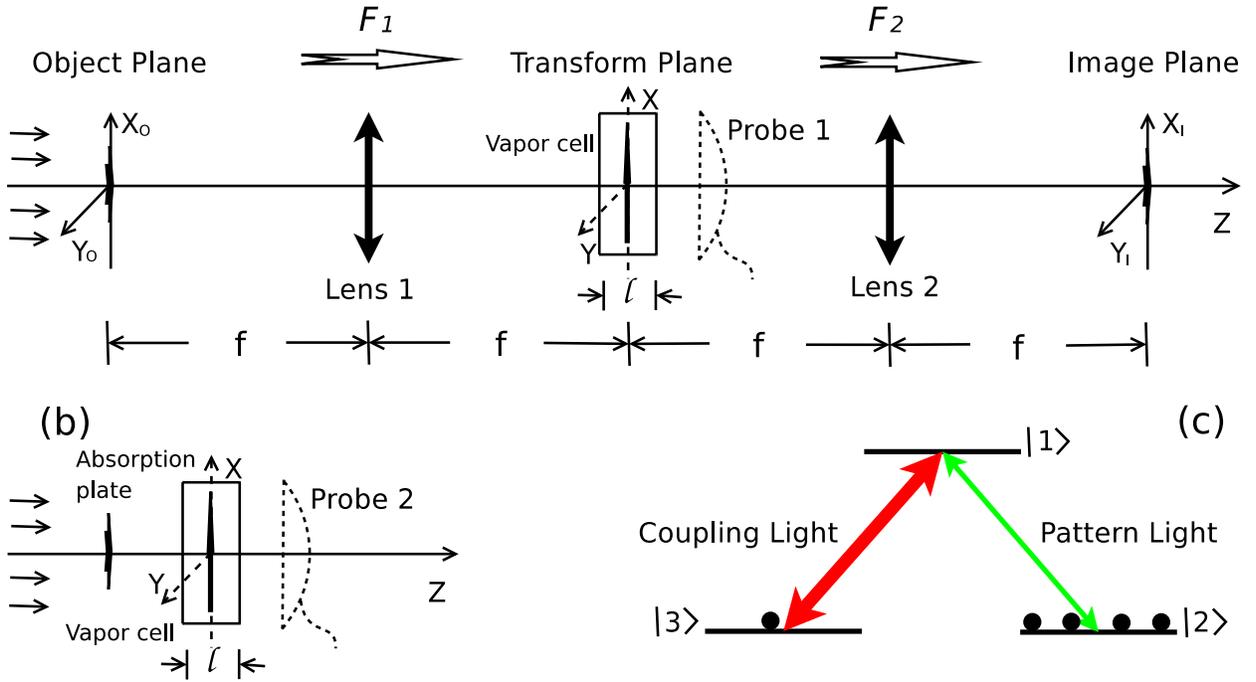}}
         \caption{(a) A diagram of a $4f$ imaging system. $F_{1}(F_{2})$ means a 
Fourier transform through lens 1 (lens 2). Probe 1 can measure the retrieved diffraction pattern. (b) A diagram for the storage of an ``artificial pattern'' in the near-field zone. Probe 2 can measure the retrieved artificial pattern. (c) The three-level $\Lambda$ system coupled with the pattern and coupling lights. For simplicity, the copropagating coupling beams and some beamsplitters are omitted.}
   \label{fig:level}
\end{figure}

        A $4f$ imaging system is shown in \fig{fig:level}a, which consists of two 
identical lenses with focal length $f$. 
An object is put in the front focal plane $(X_{O},Y_{O})$ of the first lens and the image can be retrieved in the back focal plane $(X_{I},Y_{I})$ of the second lens.  The distance between the two lenses is $2f$. The back focal plane of the first lens coincides with the front focal plane of the second lens, and is called the ``transform plane'' (TP) with coordinates $(X,Y)$ . The optical axes of the two lenses coincide, which are defined as the ``$Z$'' axis. According to Abbe theory \cite{abbe}, the Fraunhofer diffraction pattern of the object is produced near the TP. To store the pattern, a short vapor cell is placed over this plane. 
In the cell, the EIT medium can be described as a three-level $\Lambda$ system with both the pattern and coupling lights resonant with the respective 
optical transition $1 {\leftrightarrow} 2$, or $1\leftrightarrow 3$ (\fig{fig:level}c). 
In order to focus on the effect of diffusion, we neglect all the other effects of the system, such as lens aberrations, misalignment between the pattern and the coupling lights, etc. We will later justify the least obvious of those approximations.

To illustrate how the stored diffraction pattern evolves, we discuss the simplest case in which the object is a single slit of width $a$ centered around the $Z$ axis. It is uniformly illuminated by a normally incident weak pulse with central wavelength $\lambda$, whose spectral width lies within the EIT window generated by the coupling light. 
The complex amplitude of the Fraunhofer diffraction pattern of the single slit can be written
\begin{equation}
{E_{12}}^{(1)}(x) \;=\; C\,\frac{\sin(\alpha x)}{\alpha x}, \label{dp}
\end{equation}
where $C$ is a constant, and $\alpha{=}\pi a /(f\lambda)$ is the transverse wavenumber. Because the length of the cell $l$ is much smaller than the focal length ($l\ll f$), the distance between the trapped pattern and the first lens $d$ is very close to $f$. Thus, we ignore the quadratic phase factor $\exp[i \frac{\pi}{\lambda f} (1-\frac{f}{d}) x^2]$. The $sinc$-pattern of \eq{dp} has many zero crossings, i.e., dark spots. Two of them are at $ \alpha x{=}{\pm} \pi $, and the parts on both sides of the dark spots along the $X$ axis have a $\pi$ phase shift. 

For comparison, we also investigate an artificial pattern which has two dark spots at $ \alpha x{=}{\pm} \pi $ as well, but the parts on both sides of the dark spots along the $X$ axis are in phase with each other. The artificial pattern can be produced by passing a pulse through an absorption plate. 
In the near-field zone after the plate (\fig{fig:level}b), its complex amplitude could be  given by 
\begin{equation}
{E_{12}}^{(2)}(x) \;=\; C\,\left(\cos\frac{\alpha x}{2}\right)^2\exp(-\frac{x^2}{w^2}),\label{ap}
\end{equation}
where $w$ is the width of the pulse. Thus, an important difference between these two patterns is the spatial phase structure. In what follows, we can see how this difference affects the existence of dark spots.

After the coupling lights are switched off , the Raman coherence in the hot vapor is given by \cite{ls2} 
\begin{equation}
{\rho_{23}}(x,t=0) \;=\; -\frac{g}{\Omega_{13}}{E_{12}}^{(1,2)}(x), \label{lse}
\end{equation}
where $g$ is the atom-field coupling constant, $\Omega_{13}$ is the Rabi frequency of the strong coupling light before it is turned off, $t{=}0$ means \eq{lse} is the initial condition of the diffusion process, which then is described by
\begin{equation}
\frac{\partial{{\rho}_{23}(x,t)}}{\partial{t}} \;=\; D\,\frac{\partial^2\rho_{23}(x,t)}{\partial x^2}, \label{1d}
\end{equation}
where $D$ is the diffusion coefficient. To solve \eq{1d}), we introduce the 1D diffusion propagator $G(x,x^\prime,t)=(4 \pi D t)^{-1/2} \exp(-(x-x^\prime)^{2}/(4Dt))$, yielding
\begin{equation}
\rho_{23}(x,t) \;=\; \int_{-\infty}^{+\infty} \!\! \rho_{23}(x^\prime,t=0) \, G(x,x^\prime,t)\,dx^\prime. \label{1ds}
\end{equation}
 
\begin{figure}[ht]
\centerline{ \includegraphics[clip,width=0.8\linewidth]{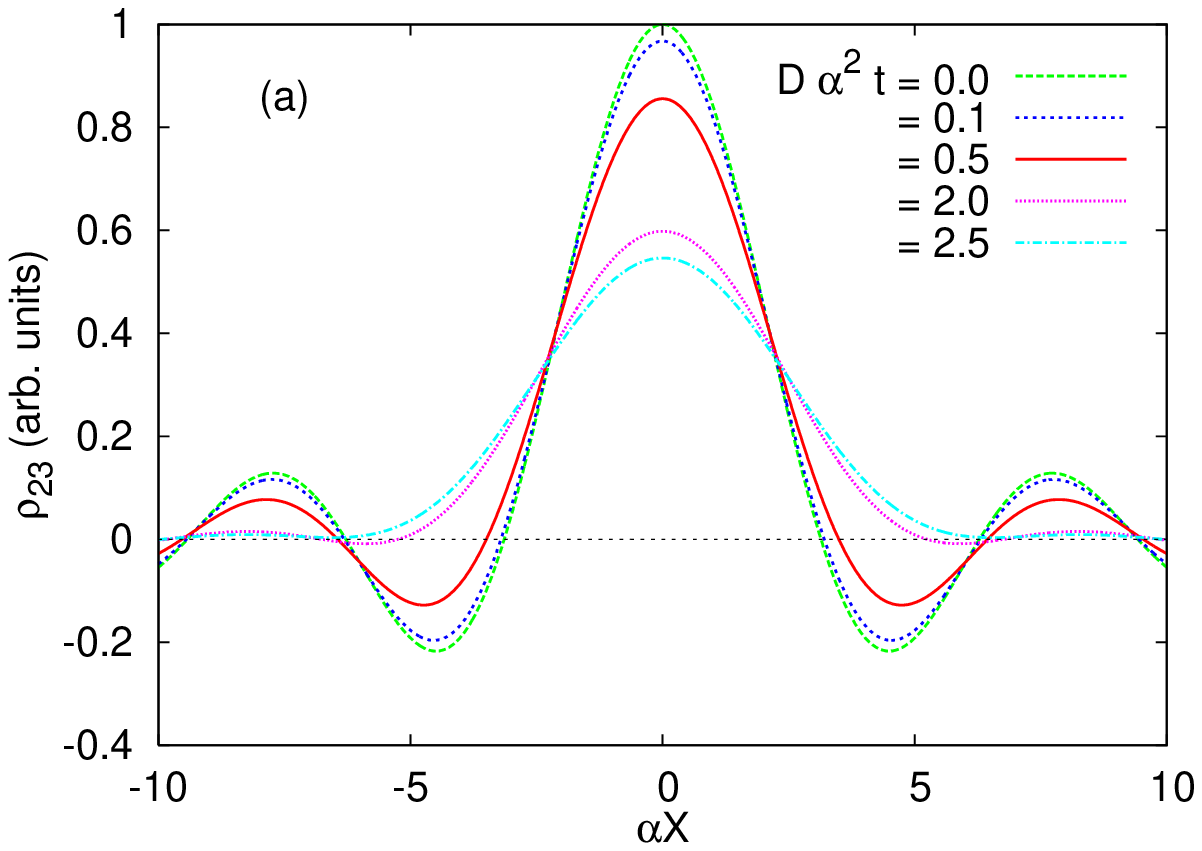}}
    \centerline{ 
        \includegraphics[clip,width=0.8\linewidth]{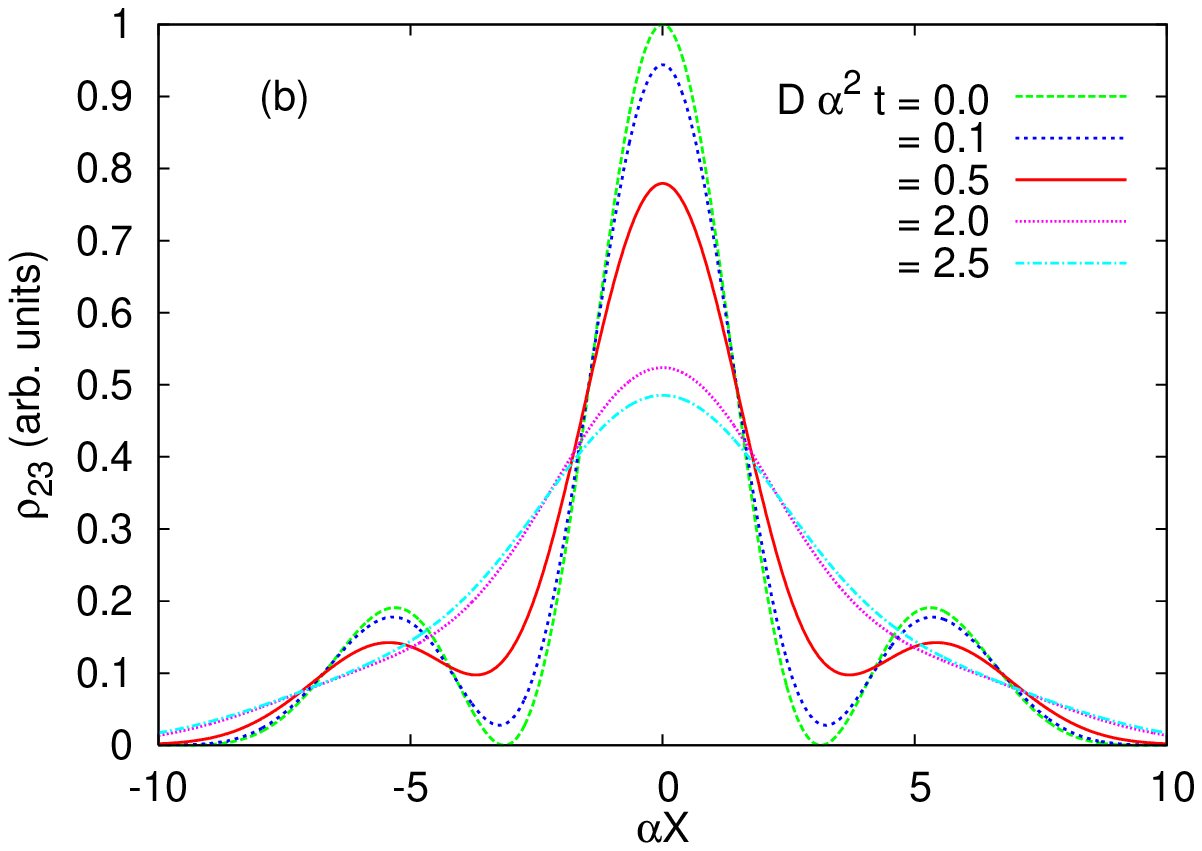}}
         \caption{(Color online) (a) The diffusion of the diffraction 
pattern. 
The dark spots will exist for a long time and move outwards. (b) The diffusion of the artificial pattern. The dark spots will disappear very quickly after the diffusion 
begins. Parameters are $f$ = 25 cm, $\lambda$ = 795  nm, $a$ = 100 $\mu$m, $w=\sqrt{20}/\alpha$. From the above graphs, we can see the dark spots of diffraction pattern still exist even when $D\alpha^2 t = 2$. From \cite{diff1}, when $D$ = 1.5 cm$^2$/s (weak diffusion), $t\approx$ 5336.5 $\mu$s; When $D$ = 30 cm$^2$/s (strong  diffusion), $t\approx$ 266.8 $\mu$s. These results show dark spots of the diffraction pattern are robust to the diffusion of atoms. All the parameters are not optimized.}
   \label{fig:raman}
\end{figure}

We can quantitatively compare the time evolution of the stored Raman coherence of  both patterns in \fig{fig:raman}. Initially, both patterns have dark spots at $\alpha x = \pm\pi$. After a very short time, the dark spots of the artificial pattern will disappear, but the dark spots of the diffraction pattern can exist for much longer and only slightly move outwards. The outward motion of the dark spots in both patterns comes from the coherence gradient of $\rho_{23}$. For the diffraction pattern which has a 1D spatial phase structure, when the out-of-phase parts diffuse into the dark spots in the opposite directions, the destructive interference will occur, and the dark spots will remain. Only until the positive phase parts cancel out all the negative phase parts, and vice versa, will the dark spots disappear. For the artificial pattern, however, everything is \emph{in} phase, no destructive interference occurs, the dark spots will be filled in right when the diffusion  beginns. 

Generally, 
the existence of dark spots in the Fraunhofer diffraction pattern can be described in terms of the destructive interference of dark-state polaritons. In principle, for some higher electromagnetic modes, such as $HG_{1,1}$  (Hermite-Gaussian) with a 2D spatial phase structure, the dark centers and lines can also be stored stably even without moving, which is due to both the destructive interference and the geometric symmetry of the phase structures. In fact, the stored optical vortex  \cite{diff2} is just a special example. Finally, when the coupling lights are switched on, the diffused patterns can be detected.

      One more point should be noted. We have only investigated 
the diffusion of Raman coherence ($\rho_{23}$) 
and seen the long-term existence of dark spots. In contrast, the atomic populations ($\rho_{22}, \rho_{33}$) are diffusing {\em without} 
interference. This can lead to fluorescence emitted in all directions in the retrieval process \cite{tun1}. 
In our probe direction (the positive $Z$ direction in \fig{fig:level}a), however, the fluorescence is very weak compared with the retrieved diffraction pattern, and will not seriously change the visibility. Thus, this effect can be neglected in our case.  

Next, we consider image formation in the image plane of \fig{fig:level}a. 
The above single slit in the $4f$ imaging system can be replaced by an actual two-dimensional object. The complex amplitude of the outgoing wave right after the object is $E_{O}(x_{O},y_{O})$. Through lens 1 the diffraction pattern in the TP can be expressed as $E(x,y) = F_{1}\left[ E_{O}(x_{O},y_{O})\right]$. While being trapped, the two-dimensional diffraction pattern will be transformed into the Raman coherence, whose motion can be described by the two-dimensional diffusion equation 
\begin{equation}
\frac{\partial\rho_{23}(x,y,t)}{\partial t} \;=\; D\,\left( \frac{\partial^2}{\partial x^2} + \frac{\partial^2}{\partial y^2}\right) \, \rho_{23}(x,y,t). \label{2d}
\end{equation}
Similar to \eq{lse}, $\rho_{23}(x,y,t)$ can be expressed through $E(x,y,t)$. 

In order to solve \eq{2d}, the Fourier transform is performed in the TP, which is seen to be $F_{2}[{E(x,y,t)}]$. Note that this expression is already the electric field in the image plane ${E}_{I}(x_{I},y_{I},t)$. For the initial condition, the Fourier transform is $F_{2}[{{E}(x,y,t{=}0)}] {=} {E}_{I}(x_{I},y_{I},t{=}0)$.

The diffusion Eq.~(\ref{2d}) with the initial condition becomes

\begin{eqnarray}
\frac{d}{dt}E_{I}(x_{I},y_{I},t) + \boldsymbol{\beta} E_{I}(x_{I},y_{I},t) &=& 0,
\label{f2d}
\end{eqnarray}
where
\[
E_{I}(x_{I},y_{I},t=0)\;=\; F_{2}[F_{1}[E_{O}(x_{O},y_{O})]]
\]
and
\[
\boldsymbol{\beta} \;=\; D \, \frac{(2 \pi)^{2} (x_I^2 + y_I^2)}{\lambda^2 f^2}.
\]

This equation is the time-evolution equation of the image we are interested in. 
The solution of \eq{f2d} is
\begin{equation}
E_{I}(x_{I},y_{I},t) \;=\; E_{O}(-x_{I},-y_{I}) \exp (-\boldsymbol{\beta} t). \label{2ds}
\end{equation}
Equation \noeq{2ds} teaches us four features of the image: (i) the image is inverted with respect to the object, (ii) the phase of the image is preserved, (iii) the amplitude at the edges of the image will decay faster than that at the center, and (iv) the dark part of the image is always dark and therefore the borders are always sharp. (The intensity is given by $I_{I}(x_{I},y_{I},t) \propto \left| E_{I}(x_{I},y_{I},t) \right|^2$.)

To quantitatively characterize the evolution of images, we simply define the fidelity by 

\begin{equation}
FI(t) \;=\;
\left| \left\langle \Psi(x_{I},y_{I},t=0)\right|\left. \Psi(x_{I},y_{I},t)\right\rangle \right|^{2},
\end{equation}
with
\begin{equation}
\Psi(x_{I},y_{I},t=0) \;=\; \frac{E_I(x_I,y_I,t=0)}{ \left( \iint\!\! \left| E_I(x_I,y_I,t=0) \right|^2 dx_I \, dy_I \right)^\frac{1}{2} }, \label{ini}
\end{equation}
and
\begin{equation}
\Psi(x_{I},y_{I},t) \;=\; \Psi(x_{I},y_{I},t{=}0) \exp (- \boldsymbol{\beta} t ),\label{tev}
\end{equation}
where \eq{ini} is the normalized initial wave function, and \eq{tev} is the time-evolving wave function. 

\begin{figure}[ht]
   \centerline{ 
   \includegraphics[angle=270,clip,width=1.0\linewidth]{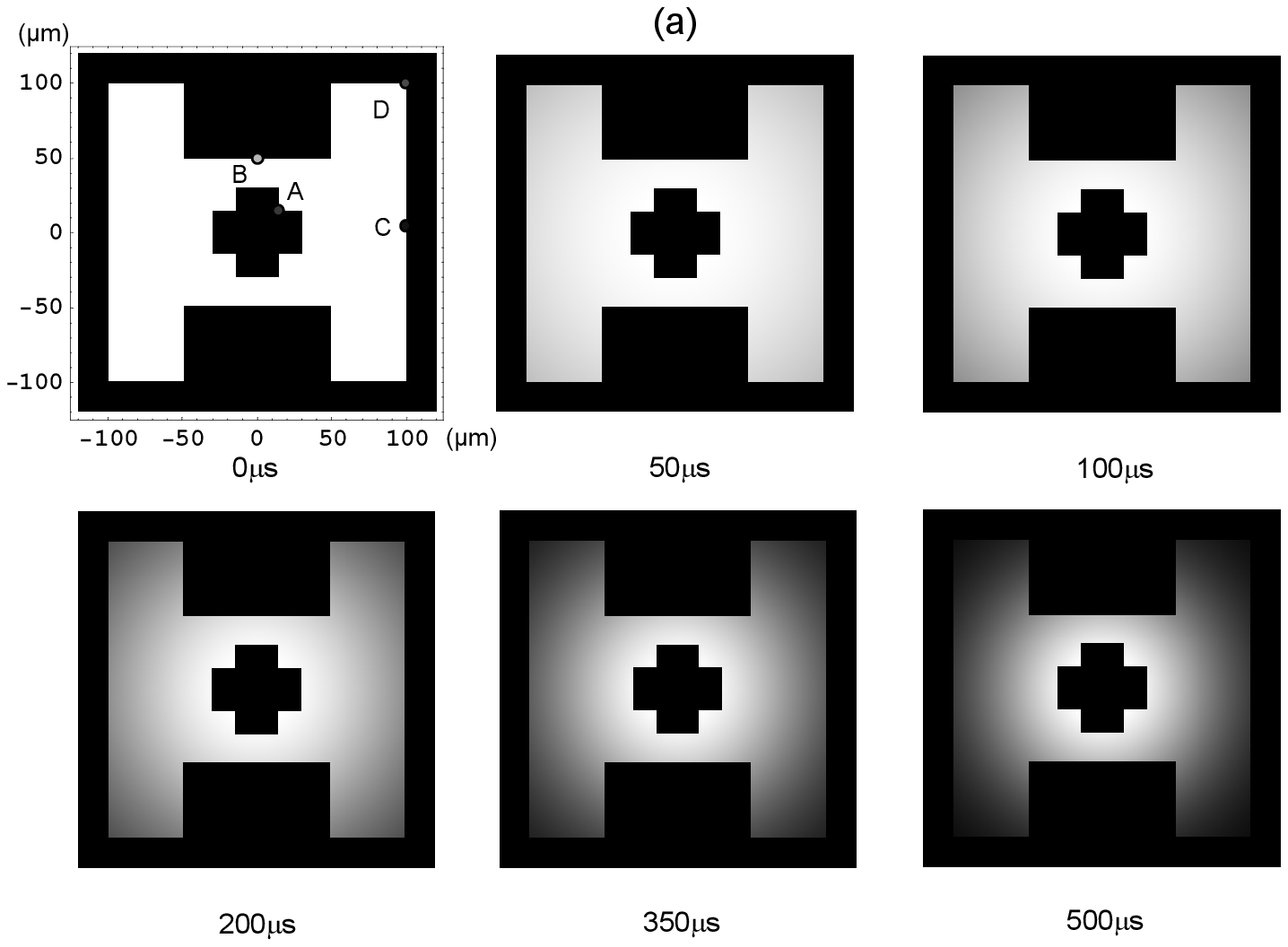}}
    \centerline{ \includegraphics[clip,width=0.5\linewidth]{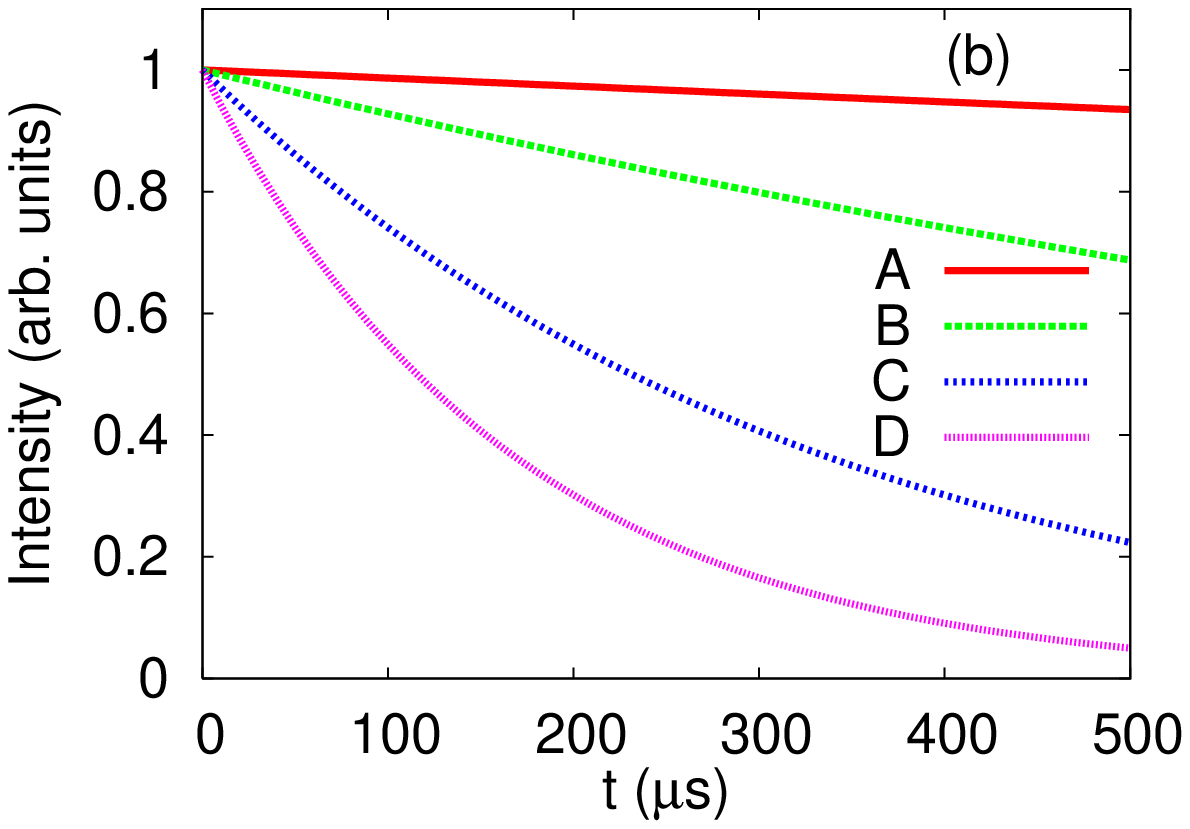} 
                 \includegraphics[clip,width=0.5\linewidth]{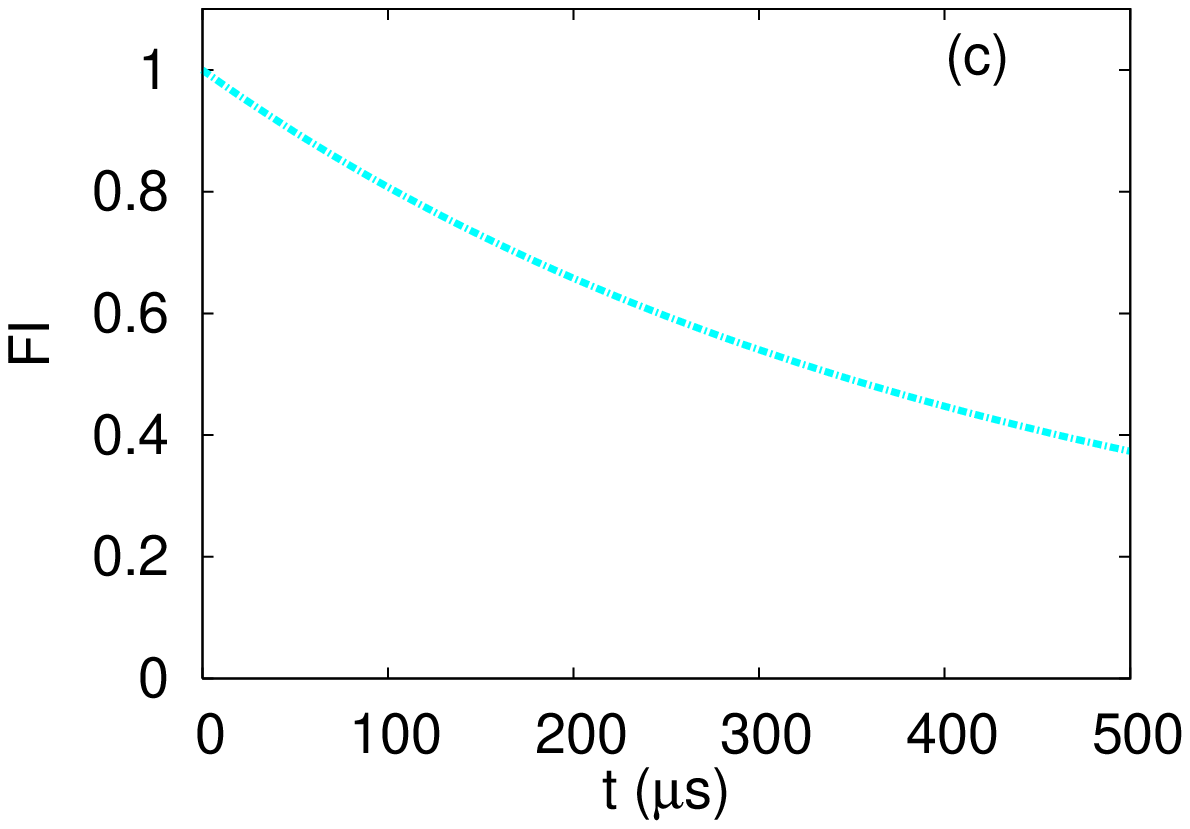}}
         \caption{(Color online) The object is centered on the $Z$ axis, 
and has a letter ``H'' with a dark cross at the center. The coordinates of points A, B, C, D are (15, 15), (0, 50), (100, 0), (100,100) in $\mu$m, respectively.  Parameters are $D$ = 1.5 cm$^2$/s, $f$ = 25 cm, $\lambda$ = 795 nm. (a) The time evolution of the image. (b) The intensities at different points varying with time.  (c) The fidelity of the whole image varying with time.  All the parameters are not optimized.}
   \label{fig:image}
\end{figure}

As an example, \fig{fig:image} shows the time evolution of the image of a letter ``H'' with a dark cross at the center, the intensities at different points in the image, and its fidelity. We can see two of above-mentioned features explicitly: (i) the binary edges of the bright image do not move, but the intensity at the edges does decrease. But, as long as the weakest intensity (e.g. the intensity at point $D$ at the corner see \fig{fig:image}b) is higher than the threshold of the detector, the edges of the image are still detectable, and the shape of the image is preserved. (ii) the dark part is always dark, no light can enter it. To explain this result, let's consider the simplest case, in which a dark wire in the object plane is illuminated by a plane wave. According to Babinet's principle \cite{babinet}, in the TP, the diffraction pattern of the dark wire is a bright spot (the diffraction pattern of the plane wave is , mathematically, a delta function) superimposed on the diffraction pattern of a single slit (the complementary object of the dark wire). Because the diffusion equation is linear, both diffraction patterns will diffuse synchronically. This can lead to the fact that the amplitude at the image of the dark wire in the image plane will always exactly cancel out and the dark part keeps dark. In fact, this phenomenon can be considered as a result of destructive interference between two diffused diffraction patterns. Generally speaking, in the whole process, if the diffraction pattern could be stored without large distortion under diffusion, it is possible to obtain a fine image. Otherwise, the image will get lost. 

Finally, we will provide some estimates of potential experimental setups. First, because the actual lenses have spherical surface, we simply estimate the aberrations \cite{optics}. Given an incident beam with the radius of 1 cm centered around and parallel to the $Z$ axis, the difference between the focal length of the margin part and that of the center part could be only about 1 mm, which is much smaller than the possible longitudinal length of the stored pattern (several cm). This means the aberrations would not seriously affect the storage of the pattern in the cell. In fact, as long as the radius of the incident beam is much smaller than that of the spherical surface, the paraxial approximation is well satisfied and the lens aberrations can be ignored. Second, because the pattern light can not be strictly parallel to the coupling light, this induces the \emph{residual} Doppler shift in hot vapors, which can lower the storage efficiency. But, in our case, the angular misalignment is small ($\sim$ 10$^{-4}$ rad) and the Dicke-like narrowing effect induced by the buffer gas collisions can strongly suppress the residual Doppler broadening \cite{dne}. Thus, our storage process is efficient. 

      In summary, we have shown that the stored dark spots of a 
Fraunhofer diffraction pattern in a $4f$ imaging system can exist for a long time  under strong diffusion. Unlike in Ref.~\cite{diff2}, the essence of such stability depends only on the destructive interference of atomic Raman coherence. Without losing generality, we show that such stability is independent of geometric dimensionality and  topological nature, 
and originates from the spatial coherence of dark-state polariton. 
Furthermore, we discuss the influence of diffusion of Raman coherence on imaging. 
In principle, this discussion could be applied to other coherent slow light or light storage processes for imaging in the media with diffusion, such as \cite{im1}. 
Although the storage of Fraunhofer diffraction pattern discussed above is a classic process, it can also go to few photon regime, 
which may have interesting applications in quantum image storage, or buffering and quantum information processing. 

We would like to acknowledge funding from NSF. L. Zhao acknowledges fruitful discussions with Prof. A. Kovner and Mr. R. Zhou.

\bibliographystyle{apsrev}
\bibliography{imagestorage}

\end{document}